\documentclass[aps,prx,superscriptaddress,reprint]{revtex4-1}
\usepackage[usenames,dvipsnames]{xcolor}
\usepackage{amsmath}
\usepackage{amssymb}
\usepackage{graphicx}
\usepackage{url}
\usepackage{hyperref}
\usepackage{color,xcolor}
\usepackage{epstopdf}
\usepackage{float}
\usepackage{ulem}
\usepackage[percent]{overpic}
\usepackage{siunitx}
\usepackage{subdepth}
\usepackage{multirow}

\usepackage{dcolumn}
\usepackage{bm}
\usepackage{times}
\usepackage[version=3]{mhchem} 

\usepackage{mathrsfs}
\usepackage{lipsum}
\usepackage{amsfonts}
\usepackage{array}
\usepackage{ulem}
\usepackage{lineno}
\usepackage[none]{hyphenat}

\hypersetup{colorlinks=true,citecolor=Red,linkcolor=Blue,urlcolor=Black}
\begin{document}
\title{Crystalline-Symmetry-Protected Majorana Modes in Coupled Quantum Dots}

\author{Bradraj Pandey}
\email{bradraj.pandey@gmail.com}
\affiliation{Department of Physics and Astronomy, The University of 
Tennessee, Knoxville, Tennessee 37996, USA}
\affiliation{Materials Science and Technology Division, Oak Ridge National 
Laboratory, Oak Ridge, Tennessee 37831, USA}

\author{Gonzalo Alvarez}
\affiliation{
Computational Sciences and Engineering Division, Oak Ridge, Tennessee 37831, USA}

\author{Elbio Dagotto}
\email{edagotto@utk.edu}
\affiliation{Department of Physics and Astronomy, The University of 
Tennessee, Knoxville, Tennessee 37996, USA}
\affiliation{Materials Science and Technology Division, Oak Ridge National 
Laboratory, Oak Ridge, Tennessee 37831, USA}

\author{Rui-Xing Zhang}
\email{ruixing@utk.edu}
\affiliation{Department of Physics and Astronomy, The University of 
Tennessee, Knoxville, Tennessee 37996, USA}
\affiliation{Department of Materials Science and Engineering, The University of 
Tennessee, Knoxville, Tennessee 37996, USA}

\date{\today}

\begin{abstract}
    We propose a minimalist architecture for achieving various crystalline-symmetry-protected Majorana modes in an array of coupled quantum dots. Our framework is motivated by the recent experimental demonstrations of two-site and three-site artificial Kitaev chains in a similar setup. We find that introducing a $\pi$-phase domain wall in the Kitaev chain leads to a pair of mirror-protected Majorana zero modes located at or near the junction. Joining two $\pi$-junctions into a closed loop, we can simulate two distinct classes of two-dimensional higher-order topological superconducting phases, both carrying symmetry-protected Majorana modes around the sample corners. As an extension of the $\pi$-junction, we further consider a general vertex structure where $n$ Kitaev chains meet, i.e., a Kitaev $n$-vertex. We prove that such an $n$-vertex, if respecting a dihedral symmetry group $D_n$, necessarily carries $n$ vertex-bound Majorana modes protected by the $D_n$ symmetry. Resilience of the junction and vertex Majorana bound states against disorder and correlation effects is also discussed. Our architecture paves the way for designing, constructing, and exploring a wide variety of artificial topological crystalline phases in experiments.       
\end{abstract}

\maketitle

\section{Introduction}

\begin{figure*}[!ht]
\hspace*{-0.5cm}
\vspace*{0cm}
\begin{overpic}[width=1.6\columnwidth]{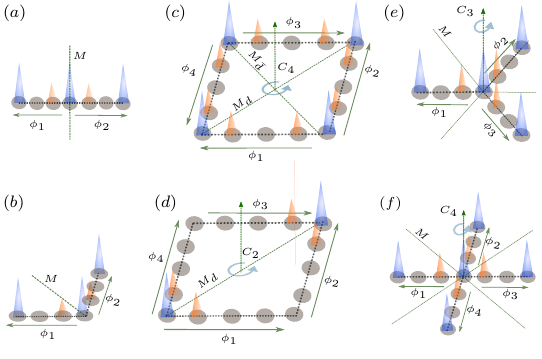}
\end{overpic}
\caption{Quantum-dot arrays with crystalline-symmetry-protected Majorana modes. (a) and (b) display two equivalent geometries of a $\pi$-phase junction with $\phi_1=\pi$ and $\phi_2=0$. The blue and orange peaks around the junction site denote two $\pi$-phase MZMs with distinct mirror parities. (c) and (d) represent two extrinsic higher-order TSC phases protected by $D_4$ and $D_2$, respectively. Here the phase configurations are  $\Phi_B=(\phi_1,\phi_2,\phi_3,\phi_4)=(\pi,0,\pi,0)$ in (c) and 
$\Phi_A=(\phi_1,\phi_2,\phi_3,\phi_4)=(0,0,\pi,\pi)$ in (d). 
  (e) shows a $D_3$-symmetric Kitaev $3$-vertex with $\phi_1=\pi$ and $\phi_2=\phi_3=0$, where one MZM and one MZD are found around the vertex. (f) shows a $D_4$-symmetric Kitaev $4$-vertex with $\phi_1=\phi_2=\pi$ and $\phi_3=\phi_4=0$, which hosts two vertex MZMs and one vertex MZD.}
\label{fig1}
\end{figure*}

Over the past decade, significant research efforts have focused on topological phenomena protected by the ubiquitous lattice symmetries in quantum crystals~\cite{LFU,Hsieh,Dziawa,LIU,Ando,Zhang1}. Starting from SnTe~\cite{LIUJ}, various topological crystalline insulators and semimetals have been theoretically predicted and experimentally revealed in a plethora of compounds, such as Na$_3$Bi~\cite{WangZ}, Cd$_3$As$_2$~\cite{Weng} , KHgSb~\cite{Cava}, MnBi$_{2n}$Te$_{3n+1}$~\cite{Zhang2}, etc. This ongoing triumph of material discovery has been greatly boosted by conceptual advances such as topological quantum chemistry~\cite{Bradlyn} and symmetry indicators~\cite{Vishwa}. Meanwhile, crystalline topological superconductors (TSCs) have similarly gathered substantial research interest~\cite{Fzhang,Teo,Sato,Shiozaki}. However, few promising real-world candidates for crystalline TSCs have been proposed and experimental confirmation remains elusive. Notably, most recipes for crystalline TSCs necessitate unconventional pairing symmetries, such as triplet superconductivity, which are rare in nature, significantly limiting the pool of potential candidates.  

Motivated by the above challenge, we propose the recently developed coupled quantum-dot system~\cite{Jay,Martin1,Martin2,Deng,Haaf} as a viable avenue to explore various crystalline topological physics of superconductors. Recently, the same platform has been exploited to realize an artificial Kitaev chain with two quantum dots (QDs) in experiments~\cite{Dvir}. We find that a $\pi$-phase junction structure of a similar QD-based Kitaev chain will trap a pair of Majorana zero modes (MZMs) protected by mirror symmetry (Figs.~\ref{fig1}(a,b)), which manifests as a building block to achieve more sophisticated crystalline topological structures. In particular, by joining a pair of $\pi$-junctions into a square, we show that we can realize two different {\it higher-order} TSC phases in two dimensions (2D) with corner-localized Majorana modes, which are protected by dihedral group symmetries $D_2$ and $D_4$, respectively (Figs.~\ref{fig1}(c,d)). 

In terms of graph theory, a $\pi$-junction can be viewed as a vertex with degree two, where each edge is a separate Kitaev chain. This further inspires us to explore the topological consequence of general vertices with $n$ Kitaev-chain edges ($n>2$), a structure dubbed ``Kitaev $n$-vertex" (Figs.~\ref{fig1}(e,f)). Remarkably, we find that an $n$-vertex, if respecting a dihedral group $D_n$, must host $n$ symmetry-indexed Majorana modes at the vertex. In particular, the vertex-bound states can be classified into singly degenerate MZMs and symmetry-enforced Majorana doublets, which exactly correspond to the 1D and 2D irreducible representations of the underlying $D_n$ group, respectively. As a proof of concept, we provide a model study of the minimal $4$-vertex with 9 QDs, which explicitly confirms the expected vertex Majorana physics. The robustness of the above crystalline-protected Majorana modes against correlation and disorder effects is also here discussed.

\section{$\pi$-Phase Junction and Mirror-Indexed Majorana Zero Modes}

Our setup of interest is a 1D chain of spin-polarized quantum dots (QDs) coupled through superconductor-semiconductor hybrids~\cite{Dvir,ChunX,Mazur, Liu2}, which is known to reproduce the famous Kitaev-chain Hamiltonian~\cite{Kitaev1},
 \begin{equation}                                                                                        H_K= \sum_{j=1}^{N-1} \left(-tc^{\dagger}_j c^{\phantom{\dagger}}_{j+1} +\Delta_jc_jc_{j+1}\right) + h.c. ,                 
 \end{equation}   
where $t$ denotes the single-electron hopping amplitude between neighboring QDs and $\Delta_j = \Delta e^{i\phi_j}$ is the nearest-neighboring triplet pairing between the $j$-th and $j+1$-th QDs. In practice, one can fine-tune the system to the ``sweet spot" with $t=\Delta$, where the topological Majorana physics  stand out. In particular, with $c_j= \frac{1}{\sqrt{2}} e^{-i\phi_j/2}\left(\gamma_{A,j}+i\gamma_{B,j}\right)$, the sweet-spot Hamiltonian becomes
\begin{equation}                                                                                        H_K = -2i\Delta \sum_{j=1}^{N-1}\gamma_{A,j+1}\gamma_{B,j}.                      
\end{equation} 
Here, the Majorana operators are anti-commutating $\{\gamma_{\alpha,j},\gamma_{\alpha',j'}\}=2\delta_{\alpha,\alpha'}\delta_{j,j'}$ and further fulfill a self-adjoint condition $\gamma_{\alpha,j}^\dagger = \gamma_{\alpha,j}$. Apparently, the end Majorana operators $\gamma_{A,1}$ and $\gamma_{B,N}$ are the only two unpaired ones that manifest as the non-Abelian Majorana zero modes (MZMs)~\cite{Kitaev1}, whose existence is $\phi_j$-independent. 

Let us now consider two QD chains of equal length and attempt to ``glue" them into a long QD chain. In the sweet-spot limit, such a gluing process is reduced to a simple two-level problem that describes the coupling of two end MZMs $\gamma_{B}^I$ and $\gamma_{A}^{II}$, where the superscript is the chain index. When two QD chains share the same pairing phase, one can always fuse and gap out the two MZMs with $H_c=-2i\Delta \gamma_{A}^{II}\gamma_{B}^I$. Using the Majorana basis $|\Psi\rangle = (|\gamma_{B,1}\rangle, |\gamma_{A,2}\rangle)^T$, $H_K=-\Delta \sigma_y$ respects both the particle-hole symmetry (PHS) $\Xi = \sigma_0 {\cal K}$ (where ${\cal K}$ is the complex conjugation operator) and a mirror symmetry $M_x = \sigma_y$ that spatially switches the dots. Note that both the spinless and odd-parity nature of the Kitaev chain requires $M_x^2=1$ and $\{M_x, \Xi\}=0$.    

Remarkably, the inter-chain gluing process is {\it impossible} when the pairings of the two QD chains differ by a $\pi$ phase. To see this, we first note that $\pi$-phase geometry implies the two QD chains have exactly the opposite pairing orders, i.e., $\Delta(x)=\Delta \text{sgn}(x)$. Here, it is helpful to use the orientation of the $p$-wave pairing to define the direction of a Kitaev chain. In this convention, the QD chains forming a $\pi$-junction are of opposite directions, as denoted by the arrows in Fig.~\ref{fig1}(a). The spatial dependence of $\Delta(x)$ not only violates the original $M_x$, but also makes the $p$-wave pairing effectively even-parity. In this case, we can always define a new mirror symmetry $\tilde{M}_x$ that is compatible with even-parity pairings, at the cost of enforcing $[\tilde{M}_x, \Xi]=0$. We then find that $\tilde{M}_x=\sigma_x$ under the Majorana basis $|\Psi\rangle$. As a consequence, any term $H_c$ that couples $\gamma_{A}^{II}$ and $\gamma_{B}^I$ must satisfy $\{\Xi, H_c\} = [\tilde{M}_x, H_c]=0$, which immediately leads to $H_c=0$. Therefore, we conclude that the $\pi$-phase domain of a 1D Kitaev TSC must bind a pair of MZMs. 

The robustness of $\pi$-phase MZMs suggests that they should be symmetry-protected. Indeed, we can combine $\gamma_{B}^I$ and $\gamma_{A}^{II}$ to form $|\gamma_{\pm} \rangle  = \frac{1}{\sqrt{2}}(|\gamma_{B}^I\rangle \pm |\gamma_{A}^{II})$. Owing to $[\tilde{M}_x, \Xi]=0$, we find that 
\begin{eqnarray}
     \tilde{M}_x |\gamma_{\pm}\rangle = \pm |\gamma_{\pm}\rangle,\ \ \Xi|\gamma_{\pm}\rangle = |\gamma_{\pm}\rangle. 
\end{eqnarray}
Therefore, $\gamma_{\pm}$ are MZMs that carry distinct $\tilde{M}_x$ indices and are hence protected by $\tilde{M}_x$. This guarantees the inability to construct a coupling $H_c$ for the $\pi$-phase MZMs, as demonstrated earlier with the algebra of Pauli matrices. 

\begin{figure}
\centering
\begin{overpic}[width=1.0\columnwidth]{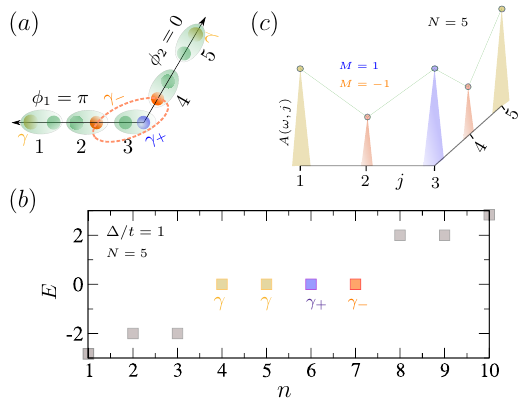}
\end{overpic}
  \caption{(a) A schematic plot of a five-QD $\pi$-junction with a pair of mirror-indexed MZMs.
  (b) BdG spectrum of the $\pi$-junction in (a) with the sweet-spot condition $\Delta/t=1$, where two edge MZMs are shown in olive. The junction MZMs $\gamma_+$ and $\gamma_-$ are shown in blue and orange, respectively.
  (c) Site-dependent local-density of states ${\cal A}(\omega=0,j)$ of the four MZMs in (b). 
  } 
\label{fig2}
\end{figure}

Let us make a few remarks. First, the existence of $\pi$-phase zero modes in Kitaev systems has been reported in the literature~\cite{Kitaev1,Roman}, but the crucial role of mirror symmetry has rarely been highlighted. By breaking $\tilde{M}_x$, accidental zero modes can also appear at the junction, but not necessarily when there is an exact $\pi$ phase difference. To summarize, the above discussion of $\pi$-phase MZMs has assumed: a) the total number of QDs $N\in$ even; b) the mirror plane lies in-between QDs. We note that an analogous proof for $N\in$ odd is straightforward when choosing the mirror plane to coincide with the location of the central QD. As a concrete example, we analytically study a QD chain with $N=5$ and further enforce a $\pi$-phase junction at $j=3$, as shown in Fig.~\ref{fig2}(a). Diagonalizing the QD Hamiltonian yields four zero-energy modes in the energy spectrum (Fig.~\ref{fig2}(b)), where two zero modes are localized around the $\pi$ junction and the other two are end MZMs~(Fig.~\ref{fig2}(c)). A further symmetry analysis reveals that the $\pi$-phase zero modes carry a mirror index of $\pm 1$, respectively, just as we expect. A detailed discussion on this $N=5$ system can be found in the Supplementary Material~\cite{SM}. Finally, while disorder effects could be inevitable in a real-world setup, we have numerically proved the resilience of the $\pi$-phase junction MZMs against moderate quench disorders that respect mirror symmetry on average~\cite{SM}. Note that similar topological robustness has also been reported in other topological crystalline systems~\cite{LiangFU}.

\section{The Poor Man's Higher-Order Topology}

The geometry of the topological QD array can be quite flexible. For example, we show in Fig.~\ref{fig1}(b) that a pair of MZMs will also emerge as {\it corner} excitations in an L-shaped junction with the two legs differing by a $\pi$ phase, as long as a leg-switching mirror symmetry is well-defined. When we assemble two identical $L$-shaped $\pi$-junctions into a square-shaped closed loop, as shown in Figs.~\ref{fig1} (c) and (d), we find two inequivalent scenarios of QD loops with two pairs of corner-localized MZMs (dubbed phase A) and four corner-MZM pairs (dubbed phase B), respectively. 

This intriguing phenomenon is reminiscent of a 2D second-order TSC, denoted as TSC$_2$, whose boundary characteristics are 0D corner-localized Majorana modes~\cite{Benalcazar,Schindler,WangLin,WangQ,YanZ,Zhang6, Zhang7,HsuY,Parames,Song,Zhang8,Zhang9}. In general, corners MZMs can be protected by either a bulk gap or an edge gap. The latter case defines an {\it extrinsic} TSC$_2$ with a nontrivial edge-state topology, whose bulk state can be topology-free~\cite{Khalaf}. In contrast, an {\it intrinsic} TSC$_2$ must have a topological ground state in the bulk. Since a closed QD loop always has a vacuum state in its interior, it is thus expected to reproduce the key physics of an extrinsic TSC$_2$ with a minimal number of degrees of freedom. We dub this strategy the ``poor man's approach"~\cite{Martin1} for TSC$_2$. 

For phase A, the SC phase for each edge is given by $\Phi_A =(\phi_1,\phi_2,\phi_3,\phi_4) = (0,0,\pi,\pi)$, as shown in Fig.~\ref{fig1} (d). We note that the system respects a dihedral symmetry group $D_2$, generated by mirror symmetries $M_d$ with $M_d(x,y)=(y,x)$ and $M_{\bar{d}}$ with $M_{\bar{d}}(x,y)=(-y,-x)$. Crucially, the phase structure $\Phi_A$ dictates $[M_d,\Xi]= \{M_{\bar{d}},\Xi\} = 0$. The commutation relations support a pair of MZM pair each $M_d$-invariant corner, but not at the $M_{\bar{d}}$-respecting corners. Notably, the only $D_2$-invariant way to eliminate the corner MZMs is to simultaneously change the topology of every Kitaev chain by closing the ``edge" gap, a manifestation of the extrinsic higher-order topology. We note that a general TSC$_2$, if being extrinsic, does not often require any symmetry protection. Nonetheless, breaking the mirror symmetries for phase A will hybridize the MZM pair at the corresponding corners, thus spoiling the second-order topology. Therefore, phase A realizes a novel special class of $D_2$-protected extrinsic TSC$_2$. 

Meanwhile, the phase structure for phase B is $\Phi_B = (\pi, 0, \pi, 0)$ (Fig.~\ref{fig1} (c)). This geometry respects a $D_4$ symmetry group generated by a four-fold rotation $C_4$ and $M_d$. Notably, both $M_d$ and $M_{\bar{d}}$ now commute with the PHS, so that both $M_d$ and $M_{\bar{d}}$-invariant corners can support MZM pairs. Such a corner-mode configuration in Fig.~\ref{fig1} (c) is also compatible with $C_4$. Phase B thus offers a poor man's version of a $D_4$-protected extrinsic TSC$_2$.  

\begin{figure}[t]
\centering
\includegraphics[width=0.48\textwidth]{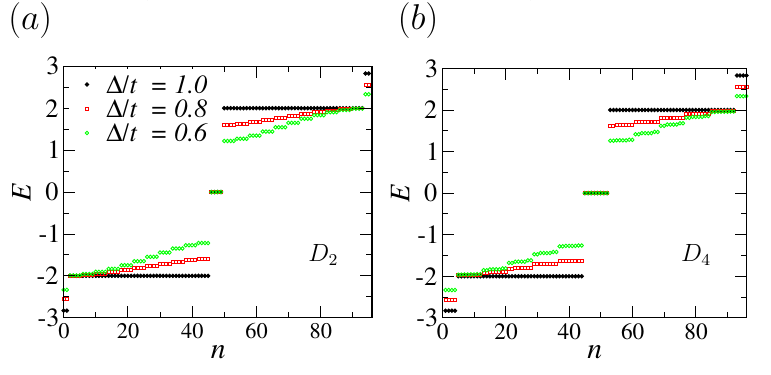}

\caption{(a) Energy spectrum of a $D_2$-symmetric higher-order TSC (Phase A), where four corner MZMs are found. (b) Energy spectrum of a $D_4$-symmetric higher-order TSC (Phase B) with eight corner MZMs. $N=48$ QDs are used in both real-space calculations. 
}
\label{fig3}
\end{figure}

Figures~\ref{fig3} (a) and (b) provide numerical simulations for both phases A and B with a total number of QDs $N=48$, which confirms the above expectations. We have also chosen different sets of model parameters, and still find the corner MZMs to be extremely localized and robust. In particular, we find that the MZM wavefunction with a positive mirror index always peaks at the corner site, while those with a negative mirror index will peak off the corner site.    

Both extrinsic TSC$_2$ phases described above share a $\mathbb{Z}_2$ classification. Namely, every mirror-respecting corner can support either zero or one pair of MZMs, if the mirror commutes with the PHS. For completeness, let us mention that an {\it intrinsic} TSC$_2$ with the same $D_2$ or $D_4$ symmetry is also $\mathbb{Z}_2$ classified, where a single MZM, rather than a MZM pair,  will show up at each mirror-respecting corner~\cite{Peng}. The intrinsic and extrinsic TSC$_2$ phases together constitute a $\mathbb{Z}_2 \times \mathbb{Z}_2$ class. This agrees with the fact that each mirror-invariant corner can host zero or one MZM in either mirror subspace.  


\section{Kitaev Vertices}

In this section, we study a new 2D geometry of QD arrays where $n$ Kitaev chains terminate at a single vertex, here is dubbed a ``Kitaev $n$-vertex". Following the language of graph theory, $n\in\mathbb{Z}_+$ is the vertex degree. Such vertex structure manifests as an elementary building block for constructing complex QD lattices, on which scalable Majorana braiding and fusion operations can be implemented~\cite{Tsintzis,Pandey1,Pandey2,JayL,Alicea}. Earlier studies have revealed the existence of zero-energy modes in certain Kitaev vertices, while the topological origin of these zero modes remains unclear~\cite{Bpandey}. As clarified below, in this paper  we show that the key to comprehending the vertex-bound zero modes is the underlying crystalline symmetry of the vertex geometry. 

\subsection{Classification of Vertex Bound States}

Let us start by labeling the edges of an $n$-vertex by an edge index $\alpha=1,2,...,n$, and an example for $n=4$ is shown in Fig.~\ref{fig1}(f). Here we have assumed $n>2$ as a $2$-vertex is essentially a linear junction structure discussed in the previous section. Since each edge is a Kitaev chain, it will contribute to an end MZM to the central node. As a result, the low-energy physics of an $n$-vertex will involve $n$ edge MZMs, dubbed $\gamma_\alpha$. Without imposing any symmetry constraint, the MZMs at a general vertex structure with an even $n$ can always be paired up and gapped out, while one vertex MZM will persist for $n=3$. Nonetheless, it is important to recall that the Kitaev chains or edges are directional. When all edges are pointing outward or inward, the $n$-vertex respects a dihedral symmetry group $D_n$, and every energy eigenstate must be labeled by an irreducible representation (irrep) of $D_n$.

\begin{table}[!t]
\begin{tabular}{c c c c} 
\hline \hline
        \ Symmetry\   & \ Degree $n$ \ & \ (${\cal N}_\text{MZM}$, ${\cal N}_{\text{MZD}}$)\  & \ ${\cal N}_\text{tot}$\  \\ \hline
\multirow{4}{*}{None} & 3 & (1,0) & 1  \\
                 &  4 & (0,0)  &  0 \\
                 & 5 & (1,0) & 1 \\
                 & 6 & (0,0) & 0 \\
                 \hline 
\multirow{4}{*}{$D_n$} & 3 & (1,1) & 3  \\
                 &  4 & (2,1)  &  4 \\
                 & 5 & (1,2) & 5 \\
                 & 6 & (2,2) & 6 \\
\hline \hline
\end{tabular}
\caption{Classification of vertex Majorana modes with and without a $D_n$ symmetry. ${\cal N}_{MZM}$ and ${\cal N}_{MZD}$ count the number of vertex-bound MZM and MZD, respectively. ${\cal N}_\text{tot}$ denotes the total number of vertex zero modes, including both MZMs and MZDs.}
\label{table1}
\end{table}

\begin{figure*}[!t]
\centering
\begin{overpic}[width=1.8\columnwidth]{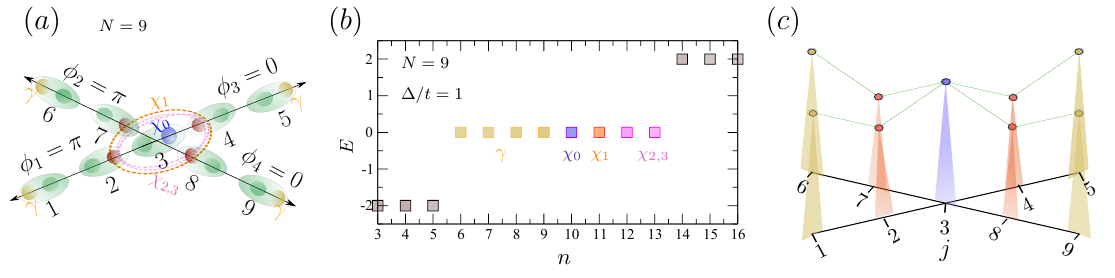}
\end{overpic}
\caption{ 
(a) A minimal $D_4$-invariant $4$-vertex with nine QDs. (b) The BdG spectrum shows eight zero modes, including four edge MZMs (in olive), two vertex MZMs (in blue and orange), and two zero modes that constitute a MZD (in magenta). (c) Real-space LDOS plots of the eight zero modes.
   }
\label{fig4}
\end{figure*}

Let us first consider the rotation $C_n$, which permutes the edge MZMs as 
\begin{equation}
    C_n\gamma_\alpha C_n^{-1} = \gamma_{\alpha+1},\ \ \text{with } \gamma_{n+1}\equiv \gamma_1.
\end{equation}
Hence, $\{\gamma_\alpha\}$ form a regular representation of $C_n$. As a consequence, the $n$ energy eigenstates from $\{\gamma_\alpha\}$ must cover all possible $|J_z\rangle$, where the $z$-component angular momentum $J_z\in\{0,1,...,n-1\}$ and $C_n|J_z\rangle = \text{exp}(i\frac{2\pi}{n}J_z)|J_z\rangle$. In the Majorana representation, the PHS $\Xi={\cal K}$ is the complex conjugation and the rotation matrix is simply a permutation matrix of different MZMs, which leads to $[C_n, \Xi]=0$. Because the PHS will flip the value of $J_z$, an $|J_z\rangle$ must be a MZM state if $J_z\equiv -J_z$ (mod $n$)~\cite{Zhang4}. It is then easy to show that
\begin{itemize}
    \item When $n$ is even, an $n$-vertex will have two vertex MZMs with $J_z=0$ and $\frac{n}{2}$. Otherwise, there will be one vertex MZM with $J_z=0$. 
\end{itemize}
Note that a similar mathematical structure of bound states has been found in the superconducting {\it vortex} of $C_n$-protected higher-order TSCs in 2D~\cite{Zhang5}.   

Meanwhile, the mirror symmetry $M$ of $D_n$ group can merge two 1D irreps into one 2D irrep. This forces $|J_z\rangle$ and $|-J_z\rangle$ to be energetically degenerate, i.e., $E_{J_z}=E_{-J_z}$. This scenario happens when $J_z\not\equiv -J_z$ (mod $n$), i.e., only when $|J_z\rangle$ is not PHS-invariant. On the other hand, $|\pm J_z\rangle$ are PHS-related, leading to $E_{J_z}=-E_{-J_z}$. We then find that
\begin{equation}
    E_{J_z} = E_{-J_z} = 0.
\end{equation}
Therefore, $|\pm J_z\rangle$ represents a new class of $D_n$-protected degenerate zero modes unveiled here that respect the PHS as a whole. We dub these new modes a ``Majorana zero doublet" (MZD), to distinguish them from the conventional non-degenerate MZMs. We have now arrived at the following remarkable conclusions:
\begin{enumerate}
    \item[(i)] A $D_n$-symmetric $n$-vertex always has $n$ zero modes, including both MZMs and MDPs;
    \item[(ii)] Vertex MZMs and MDPs are classified by the 1D and 2D irreps of $D_n$, respectively. 
\end{enumerate}

For example, a $4$-vertex should host two MZMs ($|J_z=0\rangle$ and $|J_z=2\rangle$) and one MZD contributed by $|J_z=1,3\rangle$. In total, there are four zero modes at the center of a $4$-vertex. For a $3$-vertex, there exists one MZM $|J_z=0\rangle$ and one MZD ($|J_z=1,2\rangle$) that are symmetry-protected. We can similarly identify the numbers of MZMs and MZDs for other $n$, and a summary of results can be found in Table~\ref{table1}.

\subsection{Model Study of a Minimal $4$-Vertex}

As a proof of concept, we now provide a microscopic model study for the $4$-vertex based on  QD systems. Figure~\ref{fig4} (a) shows the minimal setup for a $D_4$-symmetric $4$-vertex, where nine QDs are needed at the sweet-spot condition $\Delta=t$. As shown in Fig.~\ref{fig4} (a), $D_4$ symmetry further requires the pairing phase for each edge to be $(\phi_1, \phi_2, \phi_3, \phi_4)=(\pi,\pi,0,0)$, where the subscript denotes the edge index. Hence, a $D_4$-invariant $4$-vertex can be viewed as two $\pi$-junctions that are rotated by $\pi/2$. 

We consider the Majorana transformation defined by $c_{j}= e^{-i\phi_\alpha/2}/\sqrt{2}\left( \gamma_{A,j}+i \gamma_{B,j}\right)$, where $j=1,2,..,9$ is the site index and $\alpha$ denotes the edge index. The Hamiltonian $H^\alpha$ for each edge $\alpha$ is straightforward and can be found in the SM~\cite{SM}. The vertex Hamiltonian $H^v$ couples all the edges with the central QD at $j=3$, whose detailed form is
\begin{eqnarray}
  H^{v}=-2i\Delta \left( \gamma_{B,2} + \gamma_{B,7} + \gamma_{A,4}+  \gamma_{A,8}\right) \gamma_{B,3}.
  \label{eq:vertex coupling}
\end{eqnarray}
Interestingly, Eq.~\ref{eq:vertex coupling} implies that $\gamma_{A,3}$ is unpaired, which exactly corresponds to the vertex MZM with $J_z=0$ (denoted as $\chi_0$). Rewriting $H^v$ in the $|J_z\rangle$ basis, it is easy to show that there exist three zero-mode solutions,
\begin{eqnarray}
       \chi_1 & = & \frac{1}{2} (\gamma_{B,2} - \gamma_{B,7} + \gamma_{A,4} - \gamma_{A,8}),\nonumber \\
         \chi_2 & = & \frac{1}{2} (i\gamma_{B,2} + \gamma_{B,7} -i \gamma_{A,4} - \gamma_{A,8}),\nonumber \\ 
       \chi_3 & = & \chi_2^\dagger.
       \label{eq:4-vertex-mode}
       \end{eqnarray}
Notably, $\chi_1$ is a real MZM operator with $J_z=2$. The $J_z$ of $\chi_1$ can be read out by noting that $(\gamma_{B,2}, \gamma_{B,7}, \gamma_{A,4}, \gamma_{A,8}) \rightarrow (\gamma_{A,8}, \gamma_{B,2}, \gamma_{B,7}, \gamma_{A,4})$ under $C_4$, which implies $C_4\chi_1 C_4^{-1} = -\chi_1$. While $\chi_{2,3}$ are not self-adjoint by themselves, they carry $J_z=\pm 1$, respectively, hence corresponding to the zero-energy Majorana doublet (the MZD respect PHS as a whole~\cite{footnote}). A detailed discussion can be found in the SM~\cite{SM}. Therefore, we have analytically confirmed two MZMs ($\gamma_{A,3}$ and $\chi_1$) and one MDP ($\gamma_{2,3}$) for the minimal $4$-vertex, in full agreement with Table.~\ref{table1}. 

We have also numerically calculated the BdG spectrum of the minimal $4$-vertex with 9 QDs and found eight zero modes, as shown in Figs.~\ref{fig4} (a,b). By plotting their spatial profile in Fig.~\ref{fig4} (c),  four zero modes (highlighted in yellow) are localized at one site at the end of each edge, while $J_z=0$ MZM (colored in purple) sits at the vertex site. Notably, while the $J_z=2$ MZM (colored in orange) and the MDP (colored in pink) share exactly the same spatial location (as also shown in Eq.~\ref{eq:4-vertex-mode}), they are completely decoupled from one another thanks to the protection of $D_4$ symmetry. 

Now it should be clear why the minimal 4-vertex requires 9 QDs. If we further reduce the number of QDs (e.g., to 5), the edge MZMs would overlap with three vertex zero modes (except for the $J_z=0$ one), and no symmetry will prevent them from being hybridized. In addition, if moving away from the sweet spot, longer edges with more QDs may be needed to avoid possible hybridization between edge and vertex modes. Similarly, one can show that a minimal $3$-vertex will involve 7 QDs at the sweet-spot condition. In the SM~\cite{SM}, we have provided both the analytical and numerical results on the minimal $3$-vertex, where one MZM with $J_z=0$ and one MZD are confirmed to exist.

\begin{figure}[t]
\centering
\includegraphics[width=0.48\textwidth]{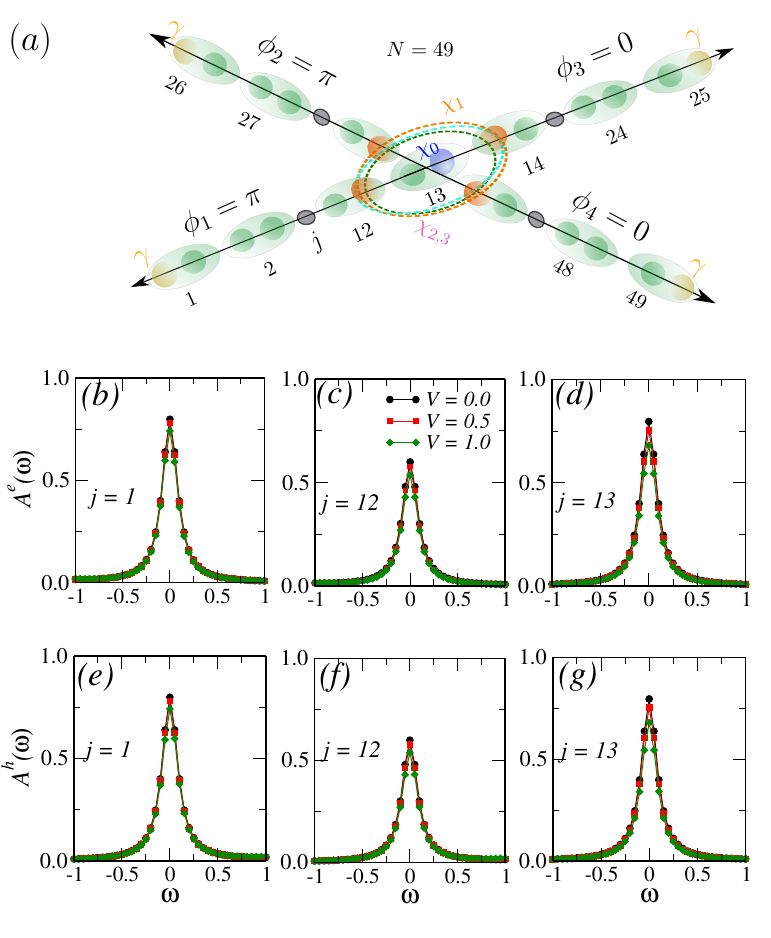}
\caption{Stability of the zero modes against a nearest-neighbor Coulomb interaction for the Kitaev $4$-vertex. (a) $N=49$ QDs are considered in the DMRG simulations and we will focus on the LDOS of lattice sites $j=1,12,13$. The electron parts of the LDOS $A^{e}(\omega)$ vs. $\omega$ 
are shown for sites (b) $j=1$ (edge site), (c) $j=12$ (near vertex), and (d) $j=13$ (vertex site). The hole part $A^{h}(\omega)$ vs. $\omega$ are shown for sites (e) $j=1$, (f) $j=12$, and (g) $j=13$. 
}
\label{fig5}
\end{figure}

Are vertex Majorana modes robust against correlation effects? To address this question, we introduce a nearest-neighbor Coulomb interaction $H_I=Vn_jn_{j+1}$ to the vertex QD system, where $n_j=c_j^\dagger c_j$ is the electron density operator at site $j$. Using the density matrix renormalization group (DMRG) method~\cite{alvarez,Nocera}, we calculate both electron and hole components of the 
local density of states (LDOS) for a $D_4$-symmetric $4$-vertex with 49 QDs, as a function of the  real frequency $\omega$ and for different $V$. Figures~\ref{fig5} (b,c,d)  show the electron LDOS ${\cal A}_e(\omega,j)$ for different sites: the edge site $j=1$, and the vertex sites $j=12$ and $j=13$, respectively. The corresponding hole LDOS for the same sites are shown in Figs.~\ref{fig5} (e,f,g). At $V=0$, the zero-bias LDOS peaks at these sites demonstrate the existence of one-site localized zero modes. In particular, we note that the zero-bias peak heights fulfill
\begin{equation}
    {\cal A}_e(0,12) = \frac{3}{4}{\cal A}_e(0,1) = \frac{3}{4}{\cal A}_e(0,13).
    \label{eq:LDOS}
\end{equation}
This relation is fully consistent with the fact that while $j=1$ and $j=13$ each host a single MZM, $j=12$ share three zero modes $\chi_{1,2,3}$ with three other sites around the vertex. Remarkably, the LDOS relation in Eq.~\ref{eq:LDOS} also holds when a finite $V$ is turned on. Note that the small $V$-dependent peak renormalization effect is likely due to the interaction-induced delocalization of the bound states. Therefore, this DMRG simulation has unambiguously illustrated the stability of vertex Majorana physics against the electron-electron interactions.

\section{Discussions}

To summarize, we have shown that junctions and vertices of QD-based artificial Kitaev chains are natural generators of multi-fold degenerate Majorana modes that carry crystalline quantum numbers. As an elementary function component of our architecture, the $\pi$-junction, along with the associated mirror-indexed MZMs, can be experimentally achieved by applying a magnetic flux through a superconducting loop that connects the two hybrid segments of quantum-dots systems~\cite{Luna}. Structures of complex geometric patterns, such as the poor man's higher-order TSC phases and Kitaev vertices, can be assembled by stacking multiple $\pi$-junctions. Fine-tuning the system to the sweet spot, a minimal geometry of the proposed junction or vertex only requires a handful of QDs, which are accessible with state-of-the-art device fabrication techniques.  Meanwhile, signatures of junction and vertex Majorana modes can be revealed in the local and nonlocal tunneling conductance measurements~\cite{Dvir, ChunX}. Away from the sweet spot, hybridizations among edge and junction/vertex Majorana modes are expected to trigger intriguing signals that are detectable with multi-terminal nonlocal transport measurements~\cite{Luzie,Martinez}. 

The dihedral-symmetric Kitaev vertices manifest as a basic element for constructing general 2D networks of connected Kitaev chains, i.e., a vertex lattice. In particular, a regular array of $3,4,6$-vertex will correspond to honeycomb, square, and triangular vertex lattices, respectively. Quasi-crystalline lattices of vertices are also possible if $5$-vertices are involved. Since each vertex is rich in Majorana degrees of freedom, turning on inter-vertex interactions in a vertex lattice opens up a new door for achieving various 2D symmetry-protected topological (SPT) states~\cite{Senthil}, topological orders~\cite{Terhal,Tessa}, fracton phases~\cite{You}, etc. We leave this intriguing directions for future discussions.

\section{Acknowledgments}
We thank S. Okamoto for helpful discussion. 
The work of B.P. and E.D. was supported by the U.S. Department of Energy, Office of Science, Basic Energy Sciences, Materials Sciences and Engineering Division. G.~A.~ was supported by the U.S. Department of Energy, Office of Science, National Quantum Information Science Research Centers, Quantum Science Center. R.-X. Z. is supported by a start-up fund of the University of Tennessee.



\begin{thebibliography}{99}
\bibitem{footnote} The 2D irrep allow us to superpose these two degenerate states so that the total PHS will be conserved.
\end{thebibliography}
\end{document}